\documentclass{article}
\usepackage{amsmath,fullpage,amssymb}
\usepackage{todonotes}
\usepackage{appendix}
\usepackage{dirtytalk}
\usepackage{authblk}
\newtheorem{rem}{Remark}
\bfseries

\begin{document}

\title{The extended versions of the noncommutative KP and mKP equations and Miura transformation}
\author{Kashif Muhammad, Li Chunxia$^{\dagger}$, Cui Mengyuan}
\affil{School of Mathematical Sciences \\
Capital Normal University\\
Beijing 100048, CHINA}

\date{}
\maketitle
\thispagestyle{empty}

\begin{abstract}
Extended versions of the noncommutative(nc) KP equation and the nc mKP equation are constructed in a unified way, for which two types of quasideterminant solutions are also presented. In commutative setting, the quasideterminant solutions provide the known and unknown Wronskian and Grammian solutions for the bilinear KP equation with self-consistent sources and the bilinear mKP equation with self-consistent sources, respectively. Miura transformation is established for the extended nc KP and nc mKP equations.
\end{abstract}\vspace{0.2cm}

{\bf Keywords:} the extended nc KP equation, the extended nc mKP equation, quasiwronskian solutions,\\ quasigrammian solutions, variation of constants, Miura transformation

\section{Introduction}

Noncommutative versions of integrable systems are of growing interest in mathematical physics. A considerable amount of literature is concerned with nc integrable systems including nc versions of the Burgers, KdV, mKdV, sine-Gorden, nonlinear Schrodinger, two-dimensional Toda lattice, KP, mKP and Davey-Stewartson equations \cite{sineGorden2002,Hamanaka_2003,2003Hamanaka_}. Generally speaking, these equations are derived by assuming that the coefficients in the Lax pair are not commutative. It has been shown that some nc integrable systems have quasideterminant solutions which can often be obtained from Darboux transformations and binary Darboux transformations \cite{2005GELFAND,2007non-Abelian,2008Li_,2007Gilson,2008Gilson,2009Li,2009Gilson_}. Besides, the quasideterminant solutions can be verified directly by using derivative formulae of quasideterminants. In this approach, it is remarkable in the sense that the results obtained are valid for super integrable systems, matrix or quaternion versions of integrable systems and Moyal-deformed integrable systems since the nature of noncommutativity is not specified and the results in commutative setting can be transformed to give the ones for the corresponding commutative integrable systems. This reveals the advantages of studying nc integrable systems.  

As an important generalization of the well-known KP equation, the nc KP equation reads as
\begin{equation*} 
    (v_t+v_{xxx}+3v_xv_x)_x-3[v_x,v_y]+3v_{yy}=0
\end{equation*}
which was proposed together with its quasiwronskian solutions and quasigrammian solutions constructed by Darboux transformations and binary Darboux transformations in \cite{2007Gilson}. In 2010, an extended nc KP hierarchy was derived by introducing square eigenfunctions in the Moyal-deformed Lax equations \cite{2010Wu}. Later on, Wu and Li made the first attempt to construct the extended nc KP hierarchy \cite{2017Li} from the existing extended KP hierarchy. Consequently, quasiwronskian solutions of the extended nc KP hierarchy were constructed by non-auto Darboux transformation, from which quasiwronskian solutions for the extended nc KP equation were obtained. However, how to construct quasigrammian solutions for the extended nc KP equation remains a problem.

As is known, the mKP equation is closely related to the KP equation through the Miura transformation. The nc mKP equation is given by \cite{2008Gilson}
\begin{equation*} 
\begin{split}
&W_x-w_y+[w,W]=0,\\
&w_t+w_{xxx}-6ww_xw+3W_y+3[w_x,W]_+-3[w_{xx},w]-3[W,w^2]=0.
\end{split}
\end{equation*}
Both quasiwronskian solutions and quasigrammian solutions were constructed for the nc mKP equation by Darboux transformation and binary Darboux transformation in \cite{2008Gilson,2009Gilson_}. An explicit connection described by Miura transformation between the quasiwronskian solutions of the nc mKP equation and the nc KP equation was verified as well. Similar to the extended nc KP equation, it is natural to ask `What does the extended nc mKP equation look like? Does it have quasiwronskian solutions and quasigrammian solutions? Is there Miura transformation between the extended nc KP equation and the extended nc mKP equation?'

In \cite{2008Wang,2005Xiao_,2009Liu}, Hu and Wang suggested the so-called source generation procedure(SGP) which is variation of constants in essence. The main idea is to introduce independent variables into arbitrary constants of the known determinant solutions or pfaffian solutions for certain soliton equations. Consequently, the newly designed determinant solutions or pfaffian solutions will satisfy new equations which are nothing but the original soliton equations with self-consistent sources(ESCSs). 
Based on this procedure, a number of soliton equations with self-consistent sources are constructed, along with which determinant solutions or pfaffian solutions are given.  Another important approach proposed by Liu etc. \cite{2005XIAO,2008LIU}  makes it possible to construct the extended KP hierarchy, the extended mKP hierarchy and so on in a unified way, which includes the KPESCS and mKPESCS as members of the corresponding hierarchies.  Solutions to these extended soliton equation hierarchies are given by Darboux transformations or by dressing approach. It is significant that these extended $(2+1)$-dimensional integrable systems and their reductions give both well-known and new integrable systems.

\begin{figure} [ht]
    \centering
    \includegraphics[width=0.7\linewidth]{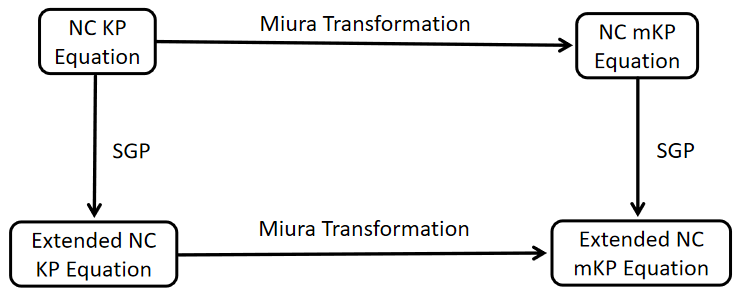}
    \caption{Miura transformation and source generation procedure}
    \label{ncKPtoncmKP}
\end{figure}

In this paper, we will first apply the source generation procedure to the nc KP equation and the nc mKP equation to generate the extended nc KP equation and the extended nc mKP equation from their corresponding quasiwronskian solutions and quasigrammian solutions, individually. Next, inspired by the Miura transformation between the nc KP equation and the nc mKP equation, we establish the Miura transformation between the extended nc KP equation and the extended nc mKP equation as shown in Figure \ref{ncKPtoncmKP}. As the benefit of studying nc integrable systems, we point out that quaswronskian solutions and quasigrammian solutions for the extended nc KP and the extended nc mKP in commutative setting can be transformed to give Wronskian solutions and Grammian solutions for the bilinear KPESCS and bilinear mKPESCS.

This paper is organized as follows. We recall some elementary properties of quasideterminants and develop a useful identity in Section 2. Section 3 is devoted to constructing the extended nc KP equation together with its quasiwronskian and quasigrammian solutions. Section 4 is devoted to constructing the extended nc mKP equation along with its quasiwronskian and quasigrammian solutions. In section 5, Miura transformation is established and proved for the extended nc KP equation and the extended nc mKP equation. Conclusions and discussions are given in Section 6.

\section{Preliminaries}
In this section, we briefly review the key elementary properties of quasideterminants and derive an useful identity. One can refer to the original papers for more details \cite{2005GELFAND,1991Gelfand}.

\subsection{Quasideterminants}
An $n\times n$ matrix $A=(a_{ij})$ over a ring (non-commutative, in general) has $n^2$ quasideterminants written as $|A|_{ij}$. They are defined recursively by
\begin{equation}
|A|_{ij}=a_{ij}-r_i^j \left(A^{ij}\right)^{-1}c_j^i=\begin{vmatrix}
    A^{ij} & c_j^i \\ r_i^j & \boxed{a_{ij}}\end{vmatrix},  \quad A^{-1}= \left(|A|_{ji} ^{-1}\right)
\end{equation}
where $r_i^j$ represents the $i$th row of $A$ with the $j$th element removed, $c_j^i$ represents the $j$th column of $A$ with the $i$th element removed, $A^{ij}$ is the submatrix of $A$ obtained by removing the $i$th row and the $j$th column from $A$. Quasideterminants can also be denoted by boxing the entry about which the expansion is made. If the entries $a_{ij}$ for  $i,j\in \{1,\cdots , n \}$ in $A$ commute, then
\begin{equation} \label{quasitocommut}
|A|_{ij}=(-1)^{i+j} \frac{\mbox{det}(A)} {\mbox{det}(A^{ij})}. 
\end{equation}

\subsection{Invariance under row and column operations}
The quasideterminants of an $n \times n$ matrix have the invariance property under certain row operations involving addition and multiplication on the left applied to the matrix: 
\begin{equation}
\begin{vmatrix}  \begin{pmatrix} E&0\\F&g \end{pmatrix} \begin{pmatrix}  A&B\\C&d  \end{pmatrix}  \end{vmatrix}_{n,n} =
\begin{vmatrix}  EA& EB \\ FA+gC & FB+gd  \end{vmatrix}_{n,n} = g(d -CA^{-1}B)=g \begin{vmatrix}  A & B \\ C & d  \end{vmatrix}_{n,n}.
\end{equation}
There is analogous invariance under column operations involving multiplication on the right.

\subsection{Nc Jacobi Identity}
Similar to the Jacobi identity for determinants, there is the nc Jacobi identity for quasideterminants:
\begin{equation} \label{jacobi}
\begin{vmatrix}A& B & C\\ D& f& g \\ E &h&\boxed{i} \end{vmatrix}= \begin{vmatrix}A& C\\E &\boxed{i}\end{vmatrix} -
\begin{vmatrix}A& B \\ E &\boxed{h} \end{vmatrix}
\begin{vmatrix}A& B \\ D& \boxed{f} \end{vmatrix} ^{-1}
\begin{vmatrix}A&C\\ D&\boxed{g} \end{vmatrix}.
\end{equation}

%where $A$, $B$ and $C$ are matrices of appropriate lengths.

\subsection{Homological relations}
The row and column homological relations for quasideterminants \cite{2005GELFAND} are given by 
\begin{equation} \label{homologicalrelation}
\begin{split}
\begin{vmatrix}A& B & C\\ D& f& g \\ E & \boxed{h} & i\end{vmatrix}= \begin{vmatrix} A & B & C\\D&f& g \\ E & h &\boxed{i}\end{vmatrix} \begin{vmatrix} A & B & C\\D&f& g \\ 0&\boxed{0}&1\end{vmatrix}, \\
\begin{vmatrix}A& B & C\\ D& f& \boxed{g} \\ E & h&i\end{vmatrix}= \begin{vmatrix}A&B&0\\D&f&\boxed{0}\\E&h&1\end{vmatrix} \begin{vmatrix} A &B&C\\D&f&g\\E&h&\boxed{i}\end{vmatrix}. 
\end{split}
\end{equation}

\subsection{Quasi-Pl\"{u}cker coordinates}
Given an $(n + k)\times n$ matrix $A$, we denote $A^i$ the $i$th row of $A$, $A^I$ the submatrix of $A$ with rows in a subset $I$ of $ \{1, 2, . . . ,n + k\} $. 
Given $ i,j \in \{1, 2,\cdots ,n + k \}$ and $I$ such that $\#I=n-1$ and $j\notin I$, the right quasi-Pl\"{u}cker coordinates are then defined for any column index $s \in \{1,\cdots,n\}$ as
\begin{equation} \label{rqpc}
r^I _ {ij} (A) = \begin{vmatrix} A^I\\ A^i  \end{vmatrix}_{ns}  \begin{vmatrix} A^I\\ A^j  \end{vmatrix}_{ns} ^{-1} = - \begin{vmatrix} A^I & 0\\ A^i & \boxed{0} \\ A^j & 1 \end{vmatrix}.
\end{equation}
The left quasi-Pl\"{u}cker coordinates can be defined in the same way.

\subsection{Derivative of Quasideterminants}
Let $A$ be a square matrix of order $N$. The derivative formula for quasideterminants is given by
\begin{equation} \label{qd}
    \begin{vmatrix} A & B \\ C & \boxed{d} \end{vmatrix}'=d'-C' A^{-1}B-C A^{-1}B'+C A^{-1}A'A^{-1}B.
\end{equation}

By denoting $e_i$ the column vector of length $N$ with $1$ in the $i$th row and $0$ elsewhere, we have
\begin{equation} \label{wd}
    \begin{vmatrix} A & B \\ C & \boxed{d} \end{vmatrix}'=\begin{vmatrix} A & B \\ C' & \boxed{d'} \end{vmatrix} +\sum_{i=1} ^N \begin{vmatrix} A & e_i \\ C & \boxed{0} \end{vmatrix} \begin{vmatrix} A & B \\ (A^i)' & \boxed{(B^i)'} \end{vmatrix}.
\end{equation}
This formula is often used when $A$ is a Wronskian-like matrix.

Suppose that $A$ is a Grammian-like matrix satisfying 
\[  A'=\sum_{i=1} ^M E_i F_i   \]
where $E_i \,(F_i)$ are column (row) vectors of length $N$. Then the derivative formula (\ref{qd}) becomes
\begin{equation} \label{gd}
    \begin{vmatrix} A & B \\ C & \boxed{d} \end{vmatrix}'=d'+\begin{vmatrix} A & B \\ C' & \boxed{0} \end{vmatrix} +\begin{vmatrix} A & B' \\ C & \boxed{0} \end{vmatrix} +\sum_{i=1} ^M \begin{vmatrix} A & E_i \\ C & \boxed{0} \end{vmatrix} \begin{vmatrix} A & B \\ F_i & \boxed{0} \end{vmatrix}.
\end{equation}

\subsection{A Useful identity}
Denote 
$$\hat{b_i} =(b_i^{(0)},  b_i^{(1)} ,\cdots, b_i^{(N-1)})^T,\quad B^{(i)} =\left( b_1^{(i)},b_2^{(i)},\cdots,b_N^{(i)}\right),\quad\hat{B}=\left(B^{(0)},B^{(1)},\cdots,B^{(N-1)}\right)^T. $$ It is clear that 
$$B^{(N)} =\sum_{i=1} ^N b_i ^{(N)}  e_i ^T,\,\quad  \hat{B} = \sum_{i=1} ^N  e_i B^{(i-1)}.$$ 
With these notations, we have the following useful identity
\begin{equation} \label{factorization}
\sum_{i=1} ^N \begin{vmatrix} A & \hat{b_i} \\ C & \boxed{b_i ^{(N)}} \end{vmatrix}  \begin{vmatrix} A & e_N \\ e_i ^T & \boxed{0} \end{vmatrix}
=\begin{vmatrix} A & e_N \\ B^{(N)} & \boxed{0} \end{vmatrix} + \sum_{i=1} ^N \begin{vmatrix} A & e_i \\ C & \boxed{0} \end{vmatrix} \begin{vmatrix} A & e_N \\ B^{(i-1)} & \boxed{0} \end{vmatrix}.
\end{equation}

\section{The extended nc KP equation}
It is known that KPSCSs reads as
\cite{1989CMaPh.126..201M} \cite{2004Xiao} 
\begin{equation} \label{ekp}
\begin{split}
&(v_t+v_{xxx}+3v_xv_x)_x+3v_{yy}= -2\left( \sum_{i=1}^M q_i r_i^\dagger \right)_x,  \\
&q_{i,y}=q_{i,xx}+v_x q_i, \\
&r_{i,y} =-r_{i,xx}- v_x^\dagger r_i
\end{split}
\end{equation}
whose Lax pair is given by
\begin{equation} \label{laxncekp}
    \begin{split}
        \phi_y&=(\partial^2+v_x)(\phi), \\
        \phi_t&=-\left[ 4\partial^3+6v_x\partial +3v_{xx}+3v_y -\sum_{i=1}^Mq_i \partial ^{-1} r_i^\dagger \right](\phi)
    \end{split}
\end{equation}
where $^\dagger$ denotes the conjugate transpose. 

By assuming that the coefficients in the Lax pair are noncommutative, we can derive the extended nc KP from the compatibility condition of \eqref{laxncekp}
\begin{equation} \label{ncekp}
    \begin{split}
&(v_t+v_{xxx}+3v_xv_x)_x+3v_{yy}-3[v_x,v_y]= -2\left( \sum_{i=1}^M q_i r_i^\dagger \right)_x,  \\
&q_{i,y}=q_{i,xx}+v_x q_i, \\
&r_{i,y} =-r_{i,xx}- v_x^\dagger r_i.
    \end{split}
\end{equation}
In the case of $q_i=r_i=0$ for $i=1,\cdots,M$, the extended nc KP \eqref{ncekp} and its Lax pair \eqref{laxncekp} are reduced to nothing but the nc KP equation \cite{2007Gilson}
\begin{equation} \label{nckp}
    \begin{split}
(v_t+v_{xxx}+3v_xv_x)_x+3v_{yy}-3[v_x,v_y]=0
    \end{split}
\end{equation}
and its Lax pair
\begin{equation} \label{laxnckp}
    \begin{split}
        \phi_y&=(\partial^2+v_x)(\phi), \\
        \phi_t&=-\left[ 4\partial^3+6v_x\partial +3v_{xx}+3v_y  \right](\phi).
    \end{split}
\end{equation}

In the following subsections, we show how to construct the extended nc KP \eqref{ncekp} from the known quasiwronskian solutions and quasigrammian solutions for the nc KP equation \eqref{nckp} by virtue of variation of constants, respectively. In contrast with Hirota's bilinear method, we illustrate how to derive the existing Wronskian and Grammian solutions to the bilinear KPESCSs from the quasiwronskian and quasigrammian solutions for the extended nc KP equation.  

\subsection{Quasiwronskian solutions}
Adopting the notations in \cite{2007Gilson}, we denote   
\begin{equation} \label{qij}
    Q(i,j)= \begin{vmatrix}
\hat{\Theta} & e_{N-j}\\
\Theta^{(N+i)} & \boxed{0}
    \end{vmatrix}
\end{equation}
 with $\Theta=(\theta_1,\cdots,\theta_N)$ and $\hat{\Theta}=\left(\theta_j^{(i-1)}\right)_{i,j=1,\ldots,N}$ being the $N\times N$ wronskian matrix of $\theta_1,\cdots,\theta_N$, where $^{(k)}$ represents the $k$th $x$-derivative and $\Theta$ satisfies the dispersion relations
\begin{equation} \label{dispersiontheta}
\Theta_y=\Theta^{(2)}  , \quad \Theta_t=-4\Theta^{(3)}.
\end{equation}  
It is noticed that $Q(i,j)$ have the properties:  
\begin{equation}
Q(i,j)=\begin{cases}
        -1 \qquad \qquad  i+j+1=0, \\
        0 \qquad  \qquad (i<0\ \mbox{or} \ j<0)\ \mbox{and} \ i+j+1\neq0.
    \end{cases}
\end{equation} 

In \cite{2007Gilson}, the following quasiwronskian solutions are constructed for the nc KP equation \eqref{nckp} by Darboux transformation
\begin{equation}
v=-2Q(0,0)=-2 \begin{vmatrix} \hat{\Theta} & e_{N}\\
\Theta^{(N)} & \boxed{0} 
    \end{vmatrix}.
\end{equation}

The quasiwronskian solutions have been verified directly as well. Actually, by relabeling and rescaling the independent variables $x_1=x$, $x_2=y$ and $x_3=-4t$, we have 
\begin{equation*}
\Theta_{x_2}=\Theta^{(2)},\quad\Theta_{x_3}=\Theta^{(3)}.\end{equation*}
Introducing higher variables $x_k$ into $\Theta$ such that $\Theta_{x_k}=\Theta^{(k)}$, we have the following general derivative formula %by using \eqref{gd}
\begin{equation} \label{Qijwithoutvc}
\partial_{x_m} Q(i,j)=Q(i+m,j)-Q(i,j+m)+ \sum_{k=1}^m Q(i,k-1)Q(m-k,j). 
\end{equation}
 The nc KP equation \eqref{nckp} becomes identically zero after substituting these derivative formulas into it.

\subsubsection{Variation of Constants}
Following the source generation procedure, let $f_i$ and $g_i\,(i=1,2,\ldots,N)$ satisfy the dispersion relations
\begin{equation} \label{dispersionfg}
 f_{i,y}=f_i^{(2)}, \quad f_{i,t}=-4f_i^{(3)}, \quad g_{i,y}=g_i^{(2)}, \quad g_{i,t}=-4g_i^{(3)}.
\end{equation}
Assume that
\begin{equation} \label{theta=f+g}
\theta_i=f_i+(-1)^{i-1}g_iC_i(t)     
\end{equation}
% \cite{2017Li}
where $C_i(t)$ is dependent of $t$ and defined by
\begin{equation} \label{defccases}
    C_i(t)=\begin{cases}
        c_i(t) \qquad 1\leq i\leq M\leq N, \\
        C_i \qquad \quad otherwise. 
    \end{cases}
\end{equation}
Consequently, we have
\begin{equation}
\theta_{i,y}=\theta_i^{(2)},\quad \theta_{i,t}= -4\theta_i^{(3)}+(-1)^{i-1}g_i \dot{c} _{i} 
\end{equation}
where $\dot{c_i}$ denote the $t$-derivative of $c_i(t)$. Or equivalently, we have
\begin{equation}\Theta^{(i)}_y=\Theta^{(i+2)},\quad \Theta_t^{(i)} =-4\Theta^{(i+3)}+H^{(i)}
\end{equation}
with
$H^{(i)}=\left((-1)^{1-1}g_1 ^{(i)} \dot c_1,\,\cdots,\,(-1)^{M-1}g_M ^{(i)} \dot c_{M},\,0,\,\cdots,\,0 \right).$ % and $\hat{H}=\left( H^{(0)},\,\cdots,\,H^{(N-1)} \right)^T $ 
%It is obvious that $v=-2Q(0,0)_t$ is exactly the quasiwronskian solutions of the nc KP equation when $C_i(t)$ is independent of $t$.

Under these assumptions, 
\begin{equation}\label{v=q} 
v=-2Q(0,0)=-2 \begin{vmatrix} \hat{\Theta} & e_{N}\\
\Theta^{(N)} & \boxed{0} 
    \end{vmatrix}
\end{equation}
will no longer satisfy the nc KP equation. It is obvious that the derivative formulae of $Q(0,0)$ with respect to $x$ and $y$ are kept the same as in the case of the nc KP equation, while the $t$-derivative of $Q(0,0)$ has additional terms determined as follows: 
\begin{equation*} 
\begin{split}
Q(0,0)_t%=&\begin{vmatrix}\hat{\Theta} & e_{N}\\\Theta ^{(N)} _t  & \boxed{0}\end{vmatrix} + \sum_{i=1} ^N \begin{vmatrix}\hat{\Theta} & e_{i} \\ \Theta^{(N)} & \boxed{0}\end{vmatrix} \begin{vmatrix}\hat{\Theta} & e_{N}\\\Theta^{(i-1)} _t & \boxed{0}\end{vmatrix} \\
=&\begin{vmatrix}\hat{\Theta} & e_{N}\\ -4\Theta^{(N+3)}+H^{(N)} & \boxed{0}\end{vmatrix} +\sum_{i=1} ^N \begin{vmatrix} \hat{\Theta} & e_{i} \\ \Theta^{(N)} & \boxed{0}\end{vmatrix} \begin{vmatrix}\hat{\Theta} & e_{N}\\ -4\Theta^{(i+2)} + H^{(i-1)}& \boxed{0}\end{vmatrix} \\
=&-4[Q(3,0)-Q(0,3)+Q(0,2)Q(0,0)+Q(0,1)Q(1,0)+Q(0,0)Q(2,0)]\\
&  + \begin{vmatrix}\hat{\Theta} & e_{N} \\ H^{(N)} & \boxed{0}\end{vmatrix} +\sum_{i=1} ^N \begin{vmatrix}\hat{\Theta}  & e_{i} \\ \Theta^{(N)} & \boxed{0}\end{vmatrix} \begin{vmatrix}\hat{\Theta} & e_{N}\\ H^{(i-1)}& \boxed{0}\end{vmatrix}\\
=&-4[Q(3,0)-Q(0,3)+Q(0,2)Q(0,0)+Q(0,1)Q(1,0)+Q(0,0)Q(2,0)]\\
&  +\sum_{i=1} ^N (-1)^{(i-1)}\begin{vmatrix}  \hat{\Theta} & \hat{g_i} \\ \Theta^{(N)} & \boxed{g_i ^{(N)}} \end{vmatrix} \dot c_i \begin{vmatrix}  \hat{\Theta} & e_N \\ e_i ^T & \boxed{0} \end{vmatrix}
\end{split}
\end{equation*}
\begin{equation*}
\begin{split}
\,\quad\qquad=&-4[Q(3,0)-Q(0,3)+Q(0,2)Q(0,0)+Q(0,1)Q(1,0)+Q(0,0)Q(2,0)]\\
\, \qquad &+\sum_{i=1} ^M (-1)^{(i-1)}\begin{vmatrix}  \hat{\Theta} & \hat{g_i} \\ \Theta^{(N)} & \boxed{g_i ^{(N)}} \end{vmatrix} \dot c_i \begin{vmatrix}  \hat{\Theta} & e_N \\ e_i ^T & \boxed{0} \end{vmatrix}
\end{split} 
\end{equation*}
where $\hat{g}_i=\left(g_i^{(0)},  g_i^{(1)},\cdots, g_i^{(N-1)}\right)^T$ and the identity \eqref{factorization} is utilized.

Actually, by applying the derivative formula \eqref{wd}, it is easy to prove that  the general derivative formula for $Q(i,j)$ are given by
\begin{equation} \label{Qij_x_m}
\partial_{x_m} Q(i,j)=Q(i+m,j)-Q(i,j+m)+ \sum_{k=1}^m Q(i,k-1)Q(m-k,j) - \frac{\delta_{m,3}}{4} \sum_{l=1} ^M S(i) \dot c_l T(j)
\end{equation}
where $\delta_{i,j}$ is the Kronecker delta and
\begin{equation} \label{ST}
S(k)=\begin{vmatrix} 
\hat{\Theta} & \hat{g_i} \\ \Theta^{(N+k)} & \boxed{g_i ^{(N+k)}} \end{vmatrix}, \quad T(j)=(-1)^{(i-1)}\begin{vmatrix}
\hat{\Theta} & e_{N-j} \\ e_i ^T & \boxed{0} \end{vmatrix}.
\end{equation} 

From now on, we shall always assume that $\dot{c_i}=\beta\eta$ with $\beta=\beta(t)$ and $\eta=\eta(t)$. 
Introduce new functions 
\begin{equation} \label{qiqw}
q_i=(-1)^NS(0)\beta_i= (-1)^N\begin{vmatrix} \hat{\Theta} & \hat{g_i} \\ \Theta^{(N)} & \boxed{g_i ^{(N)}}
\end{vmatrix} \beta_i,
\end{equation}
\begin{equation} \label{riqw}
r_i^\dagger =(-1)^N \eta_iT(0)=(-1)^{N+i-1}\eta_i \begin{vmatrix}
\hat{\Theta} & e_N \\ e_i ^T & \boxed{0} \end{vmatrix},
\end{equation} 
On one hand, by substituting $v=-2Q(0,0)$ into the nc KP equation \eqref{nckp}, we have 
\begin{equation}
(v_t+v_{xxx}+3v_xv_x)_x+3v_{yy}-3[v_x,v_y]=\left(\sum_{i=1} ^M (-1)^{(i-1)}\begin{vmatrix}  \hat{\Theta} & \hat{g_i} \\ \Theta^{(N)} & \boxed{g_i ^{(N)}} \end{vmatrix} \dot c_i \begin{vmatrix}  \hat{\Theta} & e_N \\ e_i ^T & \boxed{0} \end{vmatrix}\right)_x= -2\left( \sum_{i=1}^M q_i r_i^\dagger \right)_x
\end{equation}
which completes the proof of the first equation of the extended nc KP (\ref{ncekp}). On the other hand, we have by detailed calculations that
\begin{align*}
&q_{i,x}=(-1)^N \left[ S(1)+Q(0,0)S(0) \right] \beta_i, \qquad \qquad q_{i,y}= (-1)^N \left[ S(2)+Q(0,1)S(0)+Q(0,0)S(1) \right] \beta_i, \\
&q_{i,xx}= (-1)^N [ S(2)+Q(1,0)S(0)+\left( Q(1,0)-Q(0,1)+Q(0,0)^2 \right)S(0) + Q(0,0) \left( S(1)+Q(0,0)S(0) \right)  ] \beta_i,\\ 
&r_{i,x} ^\dagger =  (-1)^N \eta_i \left[ -T(1)+T(0)Q(0,0) \right],\quad \qquad r_{i,y} ^\dagger = (-1)^N \eta_i \left[ -T(2)+T(1)Q(0,0)+T(0)Q(1,0) \right],\\
& r_{i,xx} ^\dagger = (-1)^N \eta_i \left[ T(2)-T(0)Q(0,1)-\left(T(1)-T(0)Q(0,0)\right)Q(0,0) +T(0)\left(Q(1,0)-Q(0,1)+Q(0,0)^2\right)\right].
\end{align*}
By direct substitution, the other two constraints in \eqref{ncekp}
\begin{align*}
q_{i,y}=&q_{i,xx}+v_x q_i, \\
r_{i,y} =&-r_{i,xx}-v_x^\dagger r_i
\end{align*}
are satisfied as well. 
\begin{rem}
In the case that $C_i(t)$ in \eqref{theta=f+g} is independent of $t$ for $i=1,\ldots, N$, $v=-2Q(0,0)$ is nothing but the quasiwronskian solutions for the nc KP equation \eqref{nckp}. This explains how variation of constants is applied to generate the extended nc KP equation. 
\end{rem}

\subsubsection{Comparison with the bilinear KPESCS}
In \cite{2003Shu-fang}, the nonlinear KPESCS is derived through the linear problem of the KP equation. Wronskian solutions expressed in terms of exponential functions for the bilinear KPESCS are presented as well by Hirota's method. In \cite{2008Wang}, starting from the Grammian solutions for the bilinear KP equation, the following bilinear KPESCS are constructed by variation of constants
\begin{equation} \label{bekp}
\begin{split}
&\left( D_x^4 -4D_xD_t+3D_y^2\right) \tau \cdot \tau=8\sum_{i=1} ^M \Phi_i \Psi_i, \\
&(D_y +D_x^2) \tau \cdot \Phi_i =0, \\
&(D_y +D_x^2)  \Psi_i \cdot \tau =0
\end{split}
\end{equation}
where the Hirota's bilinear operator $D$ is defined as \cite{2004hirota}
\begin{equation} \label{hirotaderivative}
D_x^m D_y^n \tau\cdot\tau = \frac{\partial^m \partial^n}{\partial a^m \partial b^n} \tau(x+a,y+b) \tau(x-a,y-b) \bigg|_{a=0,b=0}.
\end{equation}
In addition, Wronskian solutions are constructed for the nonlinear extended KP hierarchy by dressing approach in \cite{2009Liu}. However, Wronskian solutions for the bilinear KPESCS have not been strictly and clearly stated.

Following the source generation procedure, it is not difficult to prove that the bilinear KPESCS have the Wronskian solutions expressed in terms of pfaffians
\begin{equation} \label{bekpw}
\begin{split}
&\tau=pf(d_0,\cdots,d_{N-1},N,\cdots,1), \\
&\Phi_i=  pf(d_0,\cdots,d_N,N,\cdots,1,g_i) \beta_i,  \\
&\Psi_i=  \eta_i  pf(d_0,\cdots,d_{N-2},N,\cdots,\hat{i},\cdots,1)
\end{split}
\end{equation}
where $pf(d_j,i)=\theta_i^{(j)},\, pf(i,j)=pf(d_i,d_j)=0$, $(d_j,g_i)=g_i^{(j)}$ and $\theta_i$ is given by \eqref{theta=f+g} with $\partial_{x_m}f_i=f_i^{(m)}$ and $\partial_{x_m}g_i=g_i^{(m)}$. Some necessary derivative formulae are listed in Appendix A. By direct substitution, the bilinear KPESCS is transformed into determinant identities, or equivalently, reduced pfaffian identities.

Rewrite the pfaffians mentioned above as
\[
\tau=|\hat{\Theta}|,\quad \Phi_i=\begin{vmatrix}  \hat{g_i}&\hat{\Theta}   \\  g_i ^{(N)} &\Theta^{(N)} \end{vmatrix} \beta_i, \quad 
\Psi_i= \eta_i \begin{vmatrix}  
\theta_1^{(0)}&\cdots&\theta_{i-1}^{(0)}&\theta_{i+1}^{(0)}&\cdots&\theta_N^{(0)}\\
\vdots&\ddots&\vdots&\vdots&\ddots&\vdots\\
\theta_1^{(N-2)}&\cdots&\theta_{i-1}^{(N-2)}&\theta_{i+1}^{(N-2)}&\cdots&\theta_N^{(N-2)}
\end{vmatrix}. 
\]
Notice \eqref{quasitocommut}, in commutative setting, we can establish the relations between the quasiwronskian solutions ($v,q_i,r_i^\dagger$) for the extended nc KP \eqref{ncekp}  and wronskian solutions \eqref{bekpw} for the bilinear KPSCSs \eqref{bekp}:
\begin{equation*}
v=-2Q(0,0) =2 (\ln \tau)_x,
\end{equation*}
\begin{equation*} 
q_i=(-1)^N S(0) \beta_i=\frac{\Phi_i}{\tau},
\end{equation*}
\begin{equation*} 
r_i^\dagger =(-1)^N\eta_i T(0)=\frac{\Psi_i}{\tau}
\end{equation*}
which actually provides the dependent variable transformations from the nonlinear KPESCS \eqref{ekp} to the bilinear KPESCS \eqref{bekp}.

\subsection{Quasigrammian solutions}
The general quasigrammian over nc entries presented in \cite{2007Gilson} is given by 
\begin{equation} \label{RncKP}
R(i,j)=(-1)^{j}\begin{vmatrix} {\Omega} & P^{(j)\dagger}\\
\Theta^{(i)} & \boxed{0}     \end{vmatrix}
\end{equation}
where $\Omega=\Omega(\Theta,P)_{1\leq i,j \leq N}$ is the Grammian matrix defined as
\begin{equation} \label{omega}
\Omega=\Omega(\Theta,P)=C_{ij}+\int P^\dagger \Theta  dx,
    \end{equation} 
    $\Theta=(\theta_1,\cdots,\theta_N)$ satisfies the same dispersion relations as in (\ref{dispersiontheta}), while the dispersion relations for $P=(\rho_1,\cdots,\rho_N)$ are
\begin{equation} \label{dispP}
P_y=-P_{xx}, \quad P_t=-4P_{xxx},
\end{equation}
from which, it can be easily derived that 
\[
\Omega(\Theta,P)_x=P^\dagger \Theta,\quad \Omega(\Theta,P)_y=P^\dagger \Theta_x-P^\dagger_x\Theta, \quad 
\Omega(\Theta,P)_t=-4(P^\dagger\Theta_{xx}-P^\dagger_x\Theta_x+P^\dagger_{xx}\Theta).
\]

In \cite{2007Gilson}, the following quasigrammian solutions are constructed for the nc KP \eqref{nckp} by binary Darboux transformation
\begin{equation} \label{vinR}
    v=-2R(0,0).
\end{equation}
Actually, the quasigrammian solutions can be verified directly due to the general derivative formula
\begin{equation} \label{rijwithout}
\partial_{x_m} R(i,j)=R(i+m,j)-R(i,j+m)+ \sum_{k=1}^m R(i,k-1)R(m-k,j). 
\end{equation}

\subsubsection{Variation of Constants}
Following the source generation procedure, we assume that $C_{ij}$ in \eqref{omega} are no longer constants. Instead, we have  
\begin{equation} \label{cingrammian}
C_{ij}=\begin{cases}
    c_i(t) \qquad \qquad \quad 1\leq i=j \leq M\leq N, \\
    C_i  \qquad \qquad \qquad otherwise.
\end{cases}
\end{equation}
As a result, the general derivative formula for $R(i,j)$ using \eqref{gd} can be written as
\begin{equation} \label{rijderivative}
\partial_{x_m} R(i,j)=R(i+m,j)-R(i,j+m)+ \sum_{k=1}^m R(i,k-1)R(m-k,j) - \frac{\delta_{m,3}}{4} \sum _{l=1}^M U(i) \dot{c}_l V(j)
\end{equation}
where 
\begin{equation} \label{UandV}
U(k)=(-1)^i\begin{vmatrix} \Omega  & e_i \\ \Theta^{(k)} & \boxed{0}\end{vmatrix}, \quad  V(k)=(-1)^i\begin{vmatrix} \Omega  & P^{(k)^{\dagger}} \\ e_i^T & \boxed{0}\end{vmatrix}.
\end{equation}

It is obvious that the derivative formulae of $R(0,0)$ with respect to $x$ and $y$ are kept the same as in the case of the nc KP equation \cite{2007Gilson}, while the $t$ derivative of $R(0,0)$ has additional terms. Thus we have 
\begin{equation*}
v_t=-2R(0,0)_t=8[R(3,0)-R(0,3)+R(0,2)R(0,0)+R(0,1)R(1,0)+R(0,0)R(2,0)]-2\sum_{i=1}^M U(0) \dot{c} _i V(0). 
\end{equation*}
Introduce two new functions $q_i$ and $r_i$ as follows
\begin{equation} \label{qrncekp} 
q_i= U(0) \beta_i, \quad r_i^\dagger = \eta_i V(0). 
\end{equation}
By substituting $v=-2R(0,0)$ into the nc KP equation \eqref{nckp}, it is straightforward that
\begin{equation*}
(v_t+v_{xxx}+3v_xv_x)_x+3v_{yy}-3[v_x,v_y]
=-2 \left( \sum_{i=1}^M U(0) \dot{c} _i V(0) \right)_x = -2\left( \sum_{i=1}^M q_i r_i^\dagger\right)_x
\end{equation*} 
which completes the proof of the first equation of the extended nc KP equation \eqref{ncekp}. Besides, we have by detailed
calculations that 
\begin{align*}
&q_{i,x}= \left[ U(1)+R(0,0)U(0) \right] \beta_i, \qquad q_{i,y}= \left[ U(2)+R(0,0)U(1)+R(0,1)U(0) \right] \beta_i, \\
&q_{i,xx}= \left[U(2)+R(1,0)U(0)+(R(1,0)-R(0,1)+R^2(0,0))U(0)+R(0,0)(U(1)+R(0,0)U(0))\right] \beta_i, \\  &r_{i,x}^\dagger  =\eta_i  \left[ V(1)+V(0)R(0,0) \right],\qquad r_{i,y}^\dagger  =\eta_i  \left[ -V(2)+V(0)R(1,0)-V(1)R(0,0) \right],\\
&r_{i,xx}^\dagger  =\eta_i  \left[ V(2)-V(0)R(0,1)+(V(1)+V(0)R(0,0))R(0,0)+V(0)(R(1,0)-R(0,1)+R^2(0,0))\right].
\end{align*}
By direct substitution, the other two constraints in \eqref{ncekp}
\begin{align*}
q_{i,y}=&q_{i,xx}+v_x q_i, \\
r_{i,y} =&-r_{i,xx}- v_x^\dagger r_i
\end{align*}
are satisfied as well. This completes the proof of the quasigrammian solution to the extended nc KP equation.
\begin{rem}
In the case that $C_{ij}$ for $i,j=1,\dots,N$ in \eqref{cingrammian} are independent of $t$, $v=-2R(0,0)$ is nothing but the quasigrammian solutions for the nc KP equation \eqref{nckp}. This also explains how the variation of constants is applied to generate the extended nc KP equation. 
\end{rem}

\subsubsection{Comparison with the bilinear KPESCS}
The Grammian solutions for the bilinear KPSCSs are given by \cite{2008Wang}
\begin{equation} \label{bekpg}
\begin{split}
&\tau=pf(1,\cdots,N,N^*,\cdots,1^*), \\
&\Phi_i= pf (d_0^*,1,\cdots,N ,N^*,\cdots,\hat{j^*} ,\cdots,1^*) \beta_i, \\
&\Psi_i=  \eta_i pf(d_0,1,\cdots,\hat{j},\cdots,N,N^*,\cdots,1^*)
\end{split}
\end{equation}
where $pf(i,j^*)=c_{ij}(t)+\int f_i g_j dx, \, pf(d^*_0,i)=f_i, 
\, pf(d_0,j^*)=g_j, \, pf(i^*,j^*)=pf(d_0,j)=pf(d_0^*,j^*)=0 $
with functions $f_i$ and $g_j$ satisfying the dispersion relations
\[ f_{i,y} = f_i^{(2)}, \quad  f_{i,t} = f_i^{(3)},\quad g_{i,y} = -g_i^{(2)}, \quad  g_{i,t} = g_i^{(3)}. \]

Rewrite the pfaffians mentioned above as
\begin{equation*}
\tau=|\Omega|,\quad \Phi_i= (-1)^i\begin{vmatrix} \Omega  & e_i \\ \Theta^{(0)} & 0 \end{vmatrix} \beta_i ,\quad \Psi_i=(-1)^i\eta_i \begin{vmatrix} \Omega  & P^{\dagger} \\ e_i^T & 0 \end{vmatrix}. 
\end{equation*}
In commutative setting, the quasigrammian solutions ($v,q_i,r_i^\dagger$) given by  \eqref{vinR} and \eqref{qrncekp} for the extended nc KP \eqref{ncekp} and the grammian solutions ($\tau,\Phi_i,\Psi_i$) in \eqref{bekpg} for the bilinear KPESCS satisfy the same relations as in the case of quasiwronskian solutions: 
\begin{align*} 
&v=-2R(0,0)=2(\ln \tau)_x,\\
&q_i= U(0) \beta_i =\frac{\Phi_i}{\tau
},\\ 
&r_i^\dagger  =\eta_i V(0) =\frac{\Psi_i}{\tau}. 
\end{align*}

\section{The extended nc mKP equation}
The mKPESCS reads as \cite{2009Liu}:
\begin{align*} &
W_x-w_y=0,\\
&
w_t+w_{xxx}-6ww_xw+3W_y=-\sum _{i=1} ^M \left( [w,\phi_i\varphi_i^\dagger]+(\phi_i\varphi_i^\dagger)_x \right) 
\end{align*}
with the constraints
\begin{equation*}  
\begin{split}
&\phi_{iy}=\phi_{ixx}+2w\phi_{ix}, \\
&\varphi_{iy}=-\varphi_{ixx}+2w^\dagger\varphi_{ix}
\end{split}
\end{equation*}
who has the Lax pair 
\begin{equation} \label{ncmkp-lax}
\begin{split}
&\phi_y=\left(\partial_x^{2}+2w\partial \right) \phi,\\
    &\phi_t=\left(-4\partial_x^3-12w\partial_x^2-6(w_x+w^2+W)\partial_x+\sum_{i=1}^M\phi_i\partial_x^{-1} \varphi_i^\dagger\partial_x \right) \phi.
\end{split}
\end{equation}

By assuming that the coefficients in the Lax pair \eqref{ncmkp-lax} are noncommutative, we can derive the following extended nc mKP equation from the compatibility condition
\begin{align} 
&\label{ncemkp1}
W_x-w_y+[w,W]=0,\\
 &\label{ncemkp2}
w_t+w_{xxx}-6ww_xw+3W_y+3[w_x,W]_+-3[w_{xx},w]-3[W,w^2]=-\sum _{i=1} ^M \left( [w,\phi_i\varphi_i^\dagger]+(\phi_i\varphi_i^\dagger)_x \right) 
\end{align}
with the constraints
\begin{equation} \label{ncemkp3} 
\begin{split}
&\phi_{iy}=\phi_{ixx}+2w\phi_{ix}, \\
&\varphi_{iy}=-\varphi_{ixx}+2w^\dagger\varphi_{ix}.
\end{split}
\end{equation}
If we further take $\phi_i=\varphi_i=0$, the extended nc mKP equation will be reduced to the nc mKP equation
\begin{equation} \label{ncmkp}
\begin{split}
&W_x-w_y+[w,W]=0,\\
&w_t+w_{xxx}-6ww_xw+3W_y+3[w_x,W]_+-3[w_{xx},w]-3[W,w^2]=0.
\end{split}
\end{equation} 
By taking the transformations
\begin{equation}\label{wW}
w=-F_x F^{-1}, \quad W=-F_y F^{-1}
\end{equation}
or equivalently,
\begin{equation}
w=G^{-1}G_x, \quad W=G^{-1}G_y
\end{equation}
where $G=F^{-1}$, it is obvious that \eqref{ncemkp1} is automatically satisfied.

In \cite{2008Gilson} and \cite{2009Gilson_}, both quasiwronskian solutions and quasigrammian solutions are presented for the nc KP equation \eqref{ncmkp}. In the following subsections, we will apply variation of constants to the nc mKP equation to construct the extended nc mKP based on the quasiwronskian solutions and quasigrammian solutions for the nc mKP equation, seperately.

\subsection{Quasiwronskian solutions}
The quasiwronskian solution for the nc mKP equation \eqref{ncmkp} are given by \cite{2008Gilson} 
\[ 
F=\begin{vmatrix} \Theta & \boxed{0} \\ \Theta^{(1)} &  \\ \vdots  & e_N  \\ \Theta^{(N)} &   \end{vmatrix}, \quad 
G=F^{-1}=\begin{vmatrix} \hat{\Theta} & e_1 \\ \Theta^{(N)} & \boxed{0}\end{vmatrix} 
\]
where $\Theta$ satisfies the same dispersion relations as \eqref{dispersiontheta}. By introducing the notations  
\[
F(j)= \begin{vmatrix}  \Theta & \boxed{0}  \\ \vdots  & \vdots \\\Theta^{(N-j)} & 1 \\\vdots  & \vdots  \\ \Theta^{(N)} & 0    \end{vmatrix}, \quad  
G(j)= \begin{vmatrix}  \Theta &  \\ \vdots  & e_1 \\ \Theta^{(N-1)} &  \\
\Theta^{(N+j)} & \boxed{0}    \end{vmatrix},  \quad 
Q'(i,j)= \begin{vmatrix} 
\Theta^{(1)} &  \\ \vdots  & e_{N-j} \\ \Theta^{(n)} &  \\ \Theta^{(N+1+i)} & \boxed{0}   \end{vmatrix},
\]
%Where $F=F(0)$, $G=G(0)$, $Q=Q(0,0)$, $Q'=Q'(0,0)$ (where $ v=-2Q(0,0)$,  $ \hat{v}=-2Q'(0,0)$ are the solutions of nc KP equation \cite{2007Gilson}). Also $\Theta$ satisfies the same dispersion relation , 
%$Q(i,j)$ satisfy derivative formula given in \eqref{Qijwithoutvc} and 
we have the general derivative formulae 
\begin{equation} \label{fgq'derivativewithout}
\begin{split}
F(j)_{x_{k+1}}&=F \left[ Q'(k,j)-Q(0,k+j)+\sum_{i=1}^kQ(0,i-1)Q'(k-i,j) \right],  \\
G(j)_{x_{k+1}}&= \left[ Q(k,j)-Q'(0,k+j)-\sum_{i=1}^kQ(0,i-1)Q'(k-i,j) \right]G, \\
Q'(i,j)_{x_m}&=Q'(i+m,j)-Q'(i,j+m)+ \sum_{k=1}^m Q'(i,k-1)Q'(m-k,j),
\end{split}
\end{equation}
where $Q(i,j)$ is given by \eqref{qij} and $\partial_{x_m}=\partial_x^m$ with $x_1=x,\,x_2=y,\,x_3=-4t$.  
%whose derivative formulae is given by \eqref{Qijwithoutvc}. 
Besides we denote $v=-2Q(0,0)\triangleq -2Q$. It is obvious that  $ v'=-2Q'(0,0)\triangleq-2Q' $ also provides quasiwronskian solutions for the nc KP equation. Using the properties of quasideterminants and derivative formulae given above, we have
%the homological relation \eqref{homologicalrelation}, quasi-Pl\"{u}cker coordinates \eqref{rqpc}, nc Jacobi identity \eqref{jacobi}  and above derivative formulae, some useful identities can be derived 
\begin{equation} \label{identitiesncmkp}
\begin{split}
&Q'(i,j)=Q(i+1,j-1)+Q'(i,0)Q(0,j-1), \\
&FQ(0,j)=F(j+1), \qquad\qquad Q'(j,0)G=-G(j+1), \\
&F_{xx}+F_y=2FQ'_x=-F\hat{v_x}, \quad  G_{xx}-G_y=2Q_xG=-v_xG.
\end{split}
\end{equation}
The solutions of the nc mKP equation given by (\ref{wW}) can be written as 
\begin{equation} \label{wWqr}
w=F(Q-Q')G,  \quad W=-FQ'(1,0)G-FQQ'G+FQ(0,1)G.
\end{equation}
By substitution, the quasiwronskian solutions \eqref{wWqr} for the nc mKP equation \eqref{ncmkp} have been proved by direct verification \cite{2008Gilson}.

\subsubsection{Variation of Constants}
In the same way as the extended nc KP equation, we assume 
\begin{equation*} 
\theta_i=f_i+(-1)^{i-1}g_iC_i(t)     
\end{equation*}
% \cite{2017Li}
where $f_i$ and $g_i\,(i=1,2,\ldots,N)$ satisfy the dispersion relations
\begin{equation*} 
 f_{i,y}=f_i^{(2)}, \quad f_{i,t}=-4f_i^{(3)}, \quad g_{i,y}=g_i^{(2)}, \quad g_{i,t}=-4g_i^{(3)}
\end{equation*}
and $C_i(t)$ is defined by
\begin{equation*} 
    C_i(t)=\begin{cases}
        c_i(t) \qquad 1\leq i\leq M\leq N, \\
        C_i \qquad \quad otherwise. 
    \end{cases}
\end{equation*}

Consequently, in addition to the derivative formula for $Q(i,j)$ given by \eqref{Qij_x_m}, we have the following general derivative formulae 
\begin{equation} \label{fgq'derivative}
\begin{split}
F(j)_{x_{k+1}}&=F \left[ Q'(k,j)-Q(0,k+j)+\sum_{i=1}^kQ(0,i-1)Q'(k-i,j) \right] - \frac{\delta_{k,2}}{4} \sum _{i=1}^M  Z(0) \dot c_i L(j), \\
G(j)_{x_{k+1}}&= \left[ Q(k,j)-Q'(0,k+j)-\sum_{i=1}^kQ(0,i-1)Q'(k-i,j) \right]G - \frac{\delta_{k,2}}{4} \sum _{i=1}^M  S(j) \dot c_i T(N-1), \\
Q'(i,j)_{x_m}&=Q'(i+m,j)-Q'(i,j+m)+ \sum_{k=1}^m Q'(i,k-1)Q'(m-k,j) - \frac{\delta_{m,3}}{4} \sum_{l=1} ^M  S(N+1+i) \dot c_l L(j)
\end{split}
\end{equation}
where $S(k)$ and $T(k)$ are already defined in \eqref{ST}.

Denote
\begin{equation} \label{ZandL}
Z(k)=\begin{vmatrix} \Theta^{(1)} & g_i^{(1)} \\ \vdots  & \vdots \\ \Theta^{(N)} &  g_i^{(N)} \\ \Theta^{(k)} & \boxed{g_i^{(k)}} \end{vmatrix}
, \quad 
L(k)=(-1)^{(i-1)}\begin{vmatrix} e_i^T & \boxed{0} \\ \Theta^{(1)} & \\ \vdots  & e_{N-k} \\ \Theta^{(N)} &   \end{vmatrix}
\end{equation}
 we have
\begin{equation} \label{identitiesncemkp}
F S(0)=Z(0),\quad L(0) G= - T(N-1).
\end{equation}

Given functions $\phi_i$ and $\varphi_i^\dagger$ defined by 
\begin{equation} \label{qrencmkp}
\phi_i=Z(0) \beta_i =\begin{vmatrix} \Theta & \boxed{g_i} \\ \Theta^{(1)} & g_i ^{(1)} \\ \vdots &\vdots \\ \Theta^{(N)} & g_i ^{(N)}  \end{vmatrix} \beta_i , \quad
\varphi_i^\dagger =-\eta_i T(N-1)=(-1)^i\eta_i \begin{vmatrix}  \hat{\Theta} & e_1 \\ e_i^T & \boxed{0} \end{vmatrix}
\end{equation} 
we have by derivatives and properties of quasideterminants that
\begin{align*}
 &\phi_{i,x}=F Z(N+1) \beta_i, \quad 
 \phi_{i,y}= \left( FQZ(N+1) + F Z(N+2) \right) \beta_i, \\
 &\phi_{i,xx}= [ -FQ Z(N+1)+ F Z(N+2) + 2FQ' Z(N+1) ] \beta_i, \\
&\varphi_{i,x} ^\dagger =-\eta_i T(0) G,  \qquad
\varphi_{i,y} ^\dagger =-\eta_i [ T(1) G + T(0) G(1) ],  \\  
&\varphi_{i,xx} ^\dagger =-\eta_i [ -T(1) G + 2 T(0) Q G +  T(0) G(1) ].
\end{align*}
By direct substitution, we can proof that $w$ in \eqref{wWqr}, $\phi_i$ and $\varphi_i^\dagger$ in \eqref{qrencmkp} satisfy the constraints \eqref{ncemkp3} 
\begin{equation*} 
\begin{split}
&\phi_{iy}=\phi_{ixx}+2w\phi_{ix}, \\
&\varphi_{iy}=-\varphi_{ixx}+2w^\dagger\varphi_{ix}.
\end{split}
\end{equation*}

It is noted that derivatives of $w$ with respect to $x$ or $y$ are kept the same as in \cite{2008Gilson}, while the $t$-derivative of $w$ has additional terms. Using the homological relation \eqref{homologicalrelation}, the quasi-Pl\"{u}cker coordinates \eqref{rqpc} and identities given above, we have
\begin{align*}
w_t=&F \left[ { Q'(2,0)-Q(0,2)+Q(0,0)Q'(1,0)+Q(0,1)Q'(0,0) } Q'(0,0)-\{Q'(2,0)-Q(0,2)+Q(0,0)Q'(1,0) \right. \\
&+Q(0,1)Q'(0,0)\}Q(0,0)  +Q'(3,0)-Q'(0,3) +Q'(0,2)Q'(0,0) +Q'(0,1)Q'(1,0) +Q'(0,0)Q'(2,0) \\
&-Q(3,0)+Q(0,3)-Q(0,2)Q(0,0)-Q(0,1)Q(1,0)-Q(0,0)Q(2,0) -Q'(0,0)\{Q'(2,0)-Q(0,2) \\
&\left. +Q(0,0)Q'(1,0)+Q(0,1)Q'(0,0)\} +Q(0,0)\{Q'(2,0)-Q(0,2) +Q(0,0)Q'(1,0) +Q(0,1)Q'(0,0)\} \right] G \\
&+\sum_{i=1}^M \left[   F \left( Q - Q' \right)   S(0) \dot c_i T(N-1)
+Z(0) \dot c_i L(0)  \left(  Q-Q' \right) G  +F S(0) \dot c_i T(0)  G -FZ(N+1) \dot c_i L(0)  G \right]  
\end{align*}
Notice that 
\begin{equation*}
(Q-Q')G=Gw,\quad F(Q-Q')=wF
\end{equation*}
then the additional terms in $w_t$ becomes 
\begin{align*}
&\sum_{i=1}^M \left[   F \left( Q - Q' \right)   S(0) \dot c_i T(N-1)
+Z(0) \dot c_i L(0)  \left(  Q-Q' \right) G  +F S(0) \dot c_i T(0)  G -FZ(N+1) \dot c_i L(0)  G \right]\\
&\qquad=\sum_{i=1}^M \left[  wF   S(0) \dot c_i T(N-1)
+Z(0) \dot c_i L(0)  Gw  +F S(0) \dot c_i T(0)  G -FZ(N+1) \dot c_i L(0)  G  \right]\\
&\qquad=-\sum _{i=1} ^M \left([w,\phi_i\varphi^\dagger_i]+(\phi_i\varphi^\dagger_{i})_x \right).
\end{align*}
In this way, we have proved that \eqref{wWqr} and \eqref{qrencmkp} provide the quasiwronskian solutions to \eqref{ncemkp2} too.

\subsubsection{Comparison with the bilinear mKPESCS}
The bilinear form of the mKPESCS is given by
\begin{equation} \label{bemkp}
\begin{split}
&(D_y-D_x^2) \tau' \cdot \tau=0, \\
&\left( 3D_xD_y-4D_t+D_x^3 \right) \tau' \cdot \tau =4\sum_{i=1} ^M \Phi_i \Psi_i, \\
&(D_y + D_x^2) \tau \cdot \Phi_i =0, \\
&(D_y + D_x^2)  \Psi_i \cdot \tau' =0.
\end{split}
\end{equation}
It is not difficult to prove that the mKP ESC \eqref{bemkp} has the wronskian solutions expressed in terms of pfaffians
\begin{equation} \label{bemkpw}
\begin{split}
&\tau=pf(d_0,\cdots,d_{N-1},N,\cdots,1), \qquad \quad
    \tau'=pf(d_1,\cdots,d_N,N,\cdots,1), \\
&\Phi_i=pf(d_0,\cdots,d_N,N,\cdots,1,g_i) \beta_i, \,\quad
\Psi_i=\eta_i pf(d_1,\cdots,d_{N-1},N,\cdots,\hat{i},\cdots,1),
\end{split}
\end{equation}
where
$pf(d_j,i)=\theta_i^{(j)},\, pf(i,j)=pf(d_i,d_j)=0$,\, $(d_j,g_i)=g_i^{(j}$ and $\theta_i$ is given by \eqref{theta=f+g} with $\partial_{x_m}f_i=f_i^{(m)}$ and $\partial_{x_m}g_i=g_i^{(m)}$. Some necessary derivative formulae are listed in Appendix B \cite{2004hirota}. Similar results for the extended mKP hierarchy are presented using the dressing operator approach \cite{2009Liu}. 

On one hand, rewrite \eqref{bemkpw} in terms of determinants, we have 
\begin{equation}
\tau=|\hat{\Theta}|,\quad\tau'=\begin{vmatrix} \Theta^{(1)}  \\ \vdots \\ \Theta^{(N)} \end{vmatrix},\quad \Phi_i=(-1)^N\begin{vmatrix} \Theta & g_i \\ \Theta^{(1)} & g_i ^{(1)} \\ \vdots &\vdots \\ \Theta^{(N)} & g_i ^{(N)}  \end{vmatrix},\quad \Psi_i=(-1)^i \begin{vmatrix}  \hat{\Theta} & e_1 \\ e_i ^T & 0 \end{vmatrix}.
\end{equation}
On the other hand, in commutative setting, the quasiwronskian solutions for the extended nc mKP equation are related to the Wronskian solutions for the bilinear mKPESCS as follows
\begin{align*}
&w=-F_xG=(FQ-FQ')/F=Q-Q'= -\tau_x/\tau + \tau'_x/\tau' =\partial_x ( \ln \tau'/\tau),\\
& \phi_i=  \Phi_i/\tau' , \quad 
 \varphi_i ^\dagger  =\Psi_i/\tau. 
\end{align*}

\subsection{Quasigrammian solutions}
Inspired by the quasiwronskian solutions for the extended nc mKP using \eqref{wW} derived by the quasiwronskian solutions for the nc mKP equation, we are going to find out the quasigrammian solutions for the extended nc mKP equation from certain quasigrammian solutions for the nc mKP equation. Although quasigrammian solutions are constructed for the nc mKP equation by binary Darboux transformation in \cite{2009Gilson_}, we will adopt the following different quasigrammian solutions with $F$ and $G$ given by 
\begin{align}\label{qgnckp}
F=-\begin{vmatrix} \Omega' & P^{(-1)\dagger} \\ \Theta & \boxed{-I}  \end{vmatrix}=I+F(0,-1), \quad 
 G=F^{-1}=I-R(0,-1)=\begin{vmatrix} \Omega & P^{(-1)\dagger} \\ \Theta & \boxed{I}  \end{vmatrix}
\end{align}
where $P$ and $\Theta$ are row vectors of length $N$ and satisfy the same dispersion relations defined in \eqref{dispersiontheta} and \eqref{dispP} respectively. $\Omega$ is already defined in \eqref{omega} and $\Omega'$ is defined as
\begin{equation} \label{omega'omega}
\Omega'=\Omega'(\Theta,P)= C_{ij} - \int P^{(-1)\dagger} \Theta^{(1)} dx = \Omega-P^{(-1)\dagger} \Theta.   
\end{equation}

Denote
\begin{equation} \label{HandR}
F(i,j)=(-1)^j\begin{vmatrix} \Omega' & P^{(j)\dagger} \\ \Theta^{(i)} & \boxed{0}  \end{vmatrix}, \quad 
R(i,j)=(-1)^j\begin{vmatrix} \Omega & P^{(j)\dagger} \\ \Theta^{(i)} & \boxed{0}  \end{vmatrix}.
\end{equation}
Surely, $R(i,j)$ has the same derivative formula as \eqref{rijwithout} while the general derivative formula for $F(i,j)$ is 
\begin{equation} 
F(i,j)_{x_m}=F(i+m,j)-F(i,j+m)+ \sum_{i=1}^m F(i,k-2)F(m-k+1,j). 
\end{equation}
Due to the transformation \eqref{wW}, we have
\begin{equation} \label{wWqg}
w=-FR(1,-1)+F(0,0)G, \quad W=-FR(2,-1)+F(0,1)G-F(0,0)R(1,-1).
\end{equation}
In addition, what follows are some useful identities derived by using \eqref{omega'omega} 
\begin{equation} \label{fgidentity}
\begin{split}
&F(i,-1)R(0,j)=F(i,j)-R(i,j),  \\
&R(i,-1)F(0,j)=F(i,j)-R(i,j), \\
&FR(0,j)=F(0,j)  \iff  GF(0,j)=R(0,j), \\
&F(i,-1)G=R(i,-1)  \iff R(i,-1)F=F(i,-1).
\end{split}
\end{equation}
By substitution, we can prove that \eqref{wWqg} gives nothing but the quasigrammian solutions for the nc mKP equation (\ref{ncmkp}) directly.

\subsubsection{Variation of Constants}
Now we proceed towards the quasigrammian solutions for the extended nc mKP equation by variation of constants. Let  
\begin{equation} \label{tomega'omega}
\Omega'=\Omega'(\Theta,P)= C_{ij}(t) - \int P^{(-1)\dagger} \Theta^{(1)} dx = \Omega-P^{(-1)\dagger} \Theta 
\end{equation}
with $C_{ij}(t)$ satisfies \eqref{cingrammian}. Consequently, the derivative formula for $R(i,j)$ is the same as \eqref{rijderivative}, while the general derivative formula for $F(i,j)$ becomes 
\begin{equation}
F(i,j)_{x_m}=F(i+m,j)-F(i,j+m)+ \sum_{i=1}^m F(i,k-2)F(m-k+1,j) - \frac{\delta_{m,3}}{4} \sum _{l=1}^M X(i) \dot{c}_l Y(j)
\end{equation}
where $x_1=x$, $x_2=y$, $x_3=-4t$ and the notations introduced are given by
\begin{equation} \label{XY}
X(k)=(-1)^i\begin{vmatrix} \Omega'  & e_i \\ \Theta^{(k)} & \boxed{0}\end{vmatrix}, \quad Y(k)=(-1)^{i}\begin{vmatrix} \Omega'  & P^{(k)^{\dagger}} \\ e_i^T & \boxed{0}\end{vmatrix}.
\end{equation}
Other useful identities by using the properties of quasideterminants and \eqref{omega'omega} are given as follows
\begin{equation} \label{UVXYidentity}
\begin{split}
&FU(0)=X(0) \iff  GX(0)=U(0), \\
&V(-1)F=Y(-1) \iff Y(-1)G=V(-1), \\
&F(i,-1)U(0)=X(i)-U(i)=R(i,-1)X(0), \\
&V(-1)F(0,j)=(-1)^j(V(j)-Y(j))=Y(-1)R(0,j).
\end{split}
\end{equation}

Introduce two new functions $ \phi_i$ and $\varphi_i$ as 
\begin{equation} \label{qrgencmkp}
\phi_i= X(0) \beta_i =(-1)^i\begin{vmatrix} \Omega'  & e_i \\ \Theta^{(0)} & \boxed{0}\end{vmatrix}\beta_i, \quad
\varphi_i^\dagger=-\eta_i V(-1)= (-1)^{i-1}\eta_i \begin{vmatrix} \Omega  & P^{(-1)^{\dagger}} \\ e_i^T & \boxed{0}\end{vmatrix}.
\end{equation}
The derivative formulae for $\phi_i$ and $\varphi_i^\dagger$ are given by
\begin{equation} \label{phipsiderivativencemkp}
\begin{split}
&\phi_{i,x}=[FX(1)] \beta_i, \qquad 
\phi_{i,y}=[FX(2) + F(0,0)X(1)] \beta_i, \\ 
&\phi_{i,xx}= [FX(2)+2FF(1,-1)X(1)-F(0,0)X(1)] \beta_i, \\
&\varphi_{i,x}^\dagger=- \eta_i [V(0)G], \quad  \varphi_{i,y}^\dagger= - \eta_i [-V(1)R-V(0)R(1,-1)], \\
&\varphi_{i,xx}^\dagger= -\eta_i [V(1)G+2V(0)R(0,0)G-V(0)R(1,-1)].
\end{split}
\end{equation} 

Notice that $x$-derivatives or $y$-derivatives of $F$ and $G$ are the same as in the case of the nc mKP equation, while  the $t$-derivatives of $F$ and $G$ have additional terms. The two constraints \eqref{ncemkp3} are satisfied by direct substitution. The additional terms generated by variation of constants in \eqref{ncemkp2} can be simplified as 
\begin{equation} \label{finalgencmkp}
\begin{split}&\sum_{k=1} ^M [ w X(0)\dot{c_i}V(-1)  -X(0)\dot{c_i}V(-1)w +X(0)\dot{c_i}V(0)G +FX(1)\dot{c_i}V(-1) ]  \\
&\qquad\qquad=-\sum _{k=1} ^M \left([ w,\phi_i\varphi^\dagger_i]+(\phi_i\varphi^\dagger_i)_x) \right)
\end{split}
\end{equation}
which means that \eqref{wWqg} together with \eqref{qrgencmkp} satisfy \eqref{ncemkp2}.  

So far, we have finished proving that \eqref{wWqg} together with \eqref{qrgencmkp} provide the quasigrammian solutions to the extended nc mKP equation.

\subsubsection{Comparison with the bilinear mKPESCS }
The bilinear form of the mKPESCS \eqref{bemkp} has the following Grammian solutions expressed in terms of pfaffians
\begin{equation} \label{tautau'phipsigemkp}
\begin{split}
&\tau= pf( 1,\cdots,N,N^*,\cdots,1^* ),\\
%=\det \left( C_{ij} + \int f_i^{(0)} g_j^{(0)} dx \right)_{1\leq i,j \leq N},  \\
&\tau'= \tau - pf( d_{-1},d_{0}^*,1,\cdots,N,N^*,\cdots,1^* ),\\
%=\det \left( C_{ij} - \int f_i^{(1)} g_j^{(-1)}dx \right)_{1\leq i,j \leq N} \\
&\Phi_i=pf(d_0^*,1,\cdots,N,N^*,\cdots,\hat{i^*},\cdots,1^*) \beta_i, \\
&\Psi_i=-\eta_i pf(d_{-1},1,\cdots,\hat{i},\cdots,N,N^*,\cdots,1^*)
\end{split}
\end{equation} 
where the pfaffian entries are defined as
$pf(i,j^*)=c_{ij}(t)+\int f_i g_j dx, \, $
$pf(d^*_0,i)=f_i, 
\, pf(d_0,j^*)=g_j,$
$ pf(i^*,j^*)=pf(d_0,j)=pf(d_0^*,j^*)=0 $
with functions $f_i$ and $g_j$ satisfying the dispersion relations
\[ f_{i,y} = f_i^{(2)}, \quad  f_{i,t} = f_i^{(3)},\quad g_{i,y} = -g_i^{(2)}, \quad  g_{i,t} = g_i^{(3)}. \]
To make it clear, we list all the derivatives needed for proving the quasigrammian solutions for the bilinear mKPESCS equation in Appendix C. 
%Note that the dispersion relations are different in nc emKP and bilinear emKP, because of the different choice of Lax pairs in each case.

Rewrite \eqref{tautau'phipsigemkp} in terms of Grammian determinants, we have
\begin{align}
&\tau=\det \left( C_{ij} + \int f_i^{(0)} g_j^{(0)} dx \right)_{1\leq i,j \leq N},  \\
&\tau'=\det \left( C_{ij} - \int f_i^{(1)} g_j^{(-1)}dx \right)_{1\leq i,j \leq N},\\
&\Phi_i =(-1)^i\begin{vmatrix} \Omega'   \end{vmatrix} \beta_i,\quad  \Psi_i=(-1)^{i-1}\eta_i \begin{vmatrix} \Omega & P^{\dagger(-1)} \\ e_i^T & 0 \end{vmatrix}. 
\end{align}

In commutative case, the quasigrammian solutions \eqref{wWqg} and \eqref{qrgencmkp} to the extended nc mKP equation can be expressed in terms of Grammian determinants and are closely related to the Grammian solutions for the bilinear mKPESCS as follows 
\begin{align*}
&w=-FR(1,-1)+F(0,0)G =(\tau \tau'_x - \tau' \tau_x )/(\tau \tau') = \left( \ln{\tau'/\tau} \right)_x, \\
&\phi_i=\Phi_i/\tau',\quad \varphi_i ^\dagger =\Psi_i/\tau.
\end{align*}

\section{The Miura transformation }
A Miura transformation which takes us from a solution of the nc mKP equation to that of the nc KP equation can be obtained from the Gelfand–Dikii approach \cite{2006Dimakis_}. It takes the form
\begin{equation*}
-w_x-w^2+W=Fv_xG
\end{equation*}
which inspires us to find out the Miura transformation between the extended nc KP equation and the extended nc mKP equation
\begin{equation}\label{MT}
\begin{split}
&-w_x-w^2+W=Fv_xG,  \\
&\phi_i =Fq_i, \quad\varphi^\dagger_{i,x}=-r^\dagger_i G
\end{split} 
\end{equation}
where $(v,q_i,r_i)$ is a solution of the extended nc KP equation and $(w, W, F, G, \phi_i, \varphi_i)$ is the solution of the extended nc mKP equation. It is remarkable that the Miura transformation \eqref{MT} holds true for both quasiwronskian solutions and quasigrammian solutions obtained for the extended nc KP equation and the extended nc mKP equation, respectively. The transformation \eqref{MT} can be directly verified using respective formulae and identities obtained in previous sections.

\section{Conclusions and discussions}
In this paper, firstly, a useful identity is developed. Secondly, starting from the existing quasiwronskian solutions and quasigrammian solutions for the nc KP equation, we have successfully constructed the extended nc KP equation by variation of constants, respectively. Thirdly, for the nc mKP equation, we present a slightly different quasigrammian solution. In the same way, we have succeed in constructing the extended nc mKP equation from the existing quasiwronskian solutions and the quasigrammian solutions presented by us for the nc mKP equation by variation of constants. Fourthly, in literature, Wronskian solutions in terms of exponential functions and Grammian solutions are presented for the bilinear KPESCS. For completeness, we provide the general Wronskian solutions for the bilinear KPESCS, Wronskian solutions and Grammian solutions for the bilinear mKPESCS. It is remarked that the quasiwronskian and quasigrammian solutions for the extended nc KP equation and the extended nc mKP equation in commutative setting can be transformed to give Wronskian solutions and Grammian solutions for the bilinear KPESCS and the bilinear mKPESCS, separately. The beauty and clarity of quasideterminant solutions is that it makes the verification process direct and more understandable. Finally, Miura transformation between the extended nc KP and the extended nc mKP has been established successfully and proved to be true in both quasiwronskian solutions and quasigrammian solutions, which reflects the exactness of our solutions.       

We have illustrated how to construct the extended nc KP equation and the extended nc mKP equation from the nc KP equation and the nc mKP equation by using variation of constants. It provides an efficient and unified way to construct extensions of nc integrable systems. It is believed that this approach can be used to produce extensions of other nc integrable systems such as the non-Abelian Hirota-Miwa equation and the non-Abelian Toda lattice equation. Moreover, noncommutative analogues of pfaffians have been reported and applied to several nc B-type integrable systems to derive their quasi-Pfaffian solutions by C.R. Gilson. It is also interesting to explore extensions of nc B-type integrable systems.   

\section*{Acknowledgement}

The authors would like to show their heartfelt gratitude to Professor Xing-Biao Hu for his kind guidance. This work was supported by the National Natural Science Foundation of China (Grants Nos. 11971322 and 12171475). Kashif Muhammad is deeply thankful for the generous financial support provided by the Chinese Scholarship Council (CSC No. 2021SLJ008138).

\appendix
\section*{Appendix: List of necessary derivative formulae}
\subsection*{A. Wronskian solutions for the bilinear KPESCS}
Here the derivatives of $\tau$ {\it w.r.t.} $x$ or $y$ are the same as in the case of bilinear KP equation \cite{2004hirota}.
\begin{align*}
&\tau_x=pf(d_0,\cdots,d_{N-2},d_N,N,\cdots,1),\\
&\tau_{xx}=pf(d_0,\cdots,d_{N-2},d_{N+1},N,\cdots,1)+pf(d_0,\cdots,d_{N-3},d_{N-1},d_N,N,\cdots,1), \\
&\tau_{xxx}=pf(d_0,\cdots,d_{N-2},d_{N+2},N,\cdots,1)+2pf(d_0,\cdots,d_{N-3},d_{N-1},d_{N+1},N,\cdots,1) \\
&\qquad\quad+pf(d_0,\cdots,d_{N-4},d_{N-2},d_{N-1},d_N,N,\cdots,1), \\
&\tau_{xxxx}=pf(d_0,\cdots,d_{N-2},d_{N+3},N,\cdots,1)+3pf(d_0,\cdots,d_{N-3},d_{N-1},d_{N+2},N,\cdots,1)\\
&\qquad\quad\,+2pf(d_0,\cdots,d_{N-3},d_{N},d_{N+1},N,\cdots,1)+3pf(d_0,\cdots,d_{N-4},d_{N-2},d_{N-1},d_{N+1},N,\cdots,1) \\
&\qquad\quad\,+pf(d_0,\cdots,d_{N-5},d_{N-3},d_{N-2},d_{N-1},d_N,N,\cdots,1), \\
&\tau_{y}=pf(d_0,\cdots,d_{N-2},d_{N+1},N,\cdots,1)-pf(d_0,\cdots,d_{N-3},d_{N-1},d_N,N,\cdots,1), \\
&\tau_{yy}=pf(d_0,\cdots,d_{N-2},d_{N+3},N,\cdots,1)+2pf(d_0,\cdots,d_{N-3},d_{N},d_{N+1},N,\cdots,1) \\
&\qquad\,-pf(d_0,\cdots,d_{N-4},d_{N-2},d_{N-1},d_{N+1},N,\cdots,1)-pf(d_0,\cdots,d_{N-3},d_{N-1},d_{N+2},N,\cdots,1)\\
&\qquad\,+pf(d_0,\cdots,d_{N-5},d_{N-3},d_{N-2},d_{N-1},d_N,N,\cdots,1), 
\end{align*}

\begin{align*}
&\tau_t=pf(d_0,\cdots,d_{N-2},d_{N+2},N,\cdots,1)-pf(d_0,\cdots,d_{N-3},d_{N-1},d_{N+1},N,\cdots,1) \\
 &\qquad\,+pf(d_0,\cdots,d_{N-4},d_{N-2},d_{N-1},d_N,N,\cdots,1)+\sum_{i=1} ^M \dot c_i pf(d_0,\cdots,d_{N-1},N,\cdots,\hat{i},\cdots,1,g_i) \\
&\tau_{tx}=pf(d_0,\cdots,d_{N-2},d_{N+3},N,\cdots,1)-pf(d_0,\cdots,d_{N-3},d_{N},d_{N+1},N,\cdots,1) \\
 &\qquad\,-pf(d_0,\cdots,d_{N-5},d_{N-3},d_{N-2},d_{N-1},d_N,N,\cdots,1)+\sum_{i=1} ^M \dot c_i pf(d_0,\cdots,d_{N-2},d_N,N,\cdots,\hat{i},\cdots,1,g_i), \\
    &\Phi_{i,x}= pf(d_0,\cdots,d_{N-1},d_{N+1},N,\cdots,1,g_i) \beta_i    \\
   & \Phi_{i,y}= \left[ pf(d_0,\cdots,d_{N-1},d_{N+2},N,\cdots,1,g_i) - pf(d_0,\cdots,d_{N-2},d_{N},d_{N+1},N,\cdots,1,g_i) \right] \beta_i 
     \\
   & \Phi_{i,xx}= \left[ pf(d_0,\cdots,d_{N-1},d_{N+2},N,\cdots,1,g_i) + pf(d_0,\cdots,d_{N-2},d_{N},d_{N+1},N,\cdots,1,g_i) \right]  \beta_i \\
&\Psi_{i,x}=  \eta_i  pf(d_0,\cdots,d_{N- 
   3},d_{N-1},N,\cdots,\hat{i},\cdots,1)   \\
&\Psi_{i,y}= \eta_i \left[  pf(d_0,\cdots,d_{N-3},d_{N},N,\cdots,\hat{i},\cdots,1) -pf(d_0,\cdots,d_{N- 
   4},d_{N-2},d_{N-1},N,\cdots,\hat{i},\cdots,1)   \right]   \\
& \Psi_{i,xx}=  \eta_i \left[  pf(d_0,\cdots,d_{N- 
   3},d_{N},N,\cdots,\hat{i},\cdots,1) + pf(d_0,\cdots,d_{N- 
   4},d_{N-2},d_{N-1},N,\cdots,\hat{i},\cdots,1)   \right]
\end{align*}

\subsection*{B. Wronskian solutions for the bilinear mKPESCS}
Here the derivatives of $\tau$ {\it w.r.t.} $x$ and $y$ are the same as in the case of bilinear mKP equation \cite{2004hirota}.
\begin{align*}
&\tau'_x=pf(d_1,\cdots,d_{N-1},d_{N+1},N,\cdots,1), \\
&\tau'_{xx} =pf(d_1,\cdots,d_{N-1},d_{N+2},N,\cdots,1) +pf(d_1,\cdots,d_{N-2},d_{N}, d_{N+1},N, \cdots ,1), \\
&\tau'_{xxx} =pf(d_1,\cdots,d_{N-1},d_{N+3},N,\cdots,1) +2pf (d_1,\cdots ,d_{N-2}, d_{N}, d_{N+2},N, \cdots ,1) \\
&\qquad\quad+pf(d_1,\cdots,d_{N-3},d_{N-1}, d_{N}, d_{N+1},N, \cdots ,1),  \\
&\tau'_y= pf(d_1,\cdots,d_{N-1},d_{N+2},N,\cdots,1) -pf(d_1,\cdots,d_{N-2}, d_{N},d_{N+1},N, \cdots ,1), \\
&\tau'_t=pf(d_1,\cdots,d_{N-1},d_{N+3},N,\cdots,1)-pf(d_1,\cdots,d_{N-2},d_{N},d_{N+2},N,\cdots,1) \\
 &\qquad \, +pf(d_0,\cdots,d_{N-3},d_{N-1},d_{N},d_{N+1},N,\cdots,1)+\sum_{i=1} ^M \dot c_i pf(d_1,\cdots,d_{N},N,\cdots,\hat{i},\cdots,1,g_i) \\
&\Psi_{i}=  \eta_i pf(d_1,\cdots,d_{N-1},N,\cdots,\hat{i},\cdots,1) \\ 
&\Psi_{i,x}=  \eta_i  pf(d_1,\cdots,d_{N-2},d_{N},N,\cdots,\hat{i},\cdots,1) \\ 
&\Psi_{i,xx}=  \eta_i \left[ pf(d_1,\cdots,d_{N-2}, d_{N+1}, N,\cdots, \hat{i},\cdots ,1)+pf(d_1,\cdots,d_{N-3}, d_{N-1},d_{N},N, \cdots, \hat{i}, \cdots,1)   \right]  \\  
&\Psi_{i,y}= \eta_i \left[  pf(d_1,\cdots,d_{N-2}, d_{N+1},N ,\cdots,\hat{i}, \cdots,1) - pf(d_1,\cdots,d_{N-3},d_{N-1},d_{N},N, \cdots, \hat{i},\cdots,1)  \right]
\end{align*}
Here we have skipped the derivatives of $\tau$ and $\Phi$ which are the same as in Appendix A.
\subsection*{C. Grammian solutions for the bilinear mKPESCS}
Here the derivatives of $\tau$ and $\tau'$ {\it w.r.t.} $x$ and $y$ are the same as in the case of bilinear mKP equation \cite{2004hirota}.
\begin{align*}
&\tau_x=pf(d_0,d_0^*,1,\cdots,N,N^*,\cdots,1^*), \\ &\tau_{xx}=pf(d_1,d_0^*,1,\cdots,N,N^*,\cdots,1^*)+pf(d_0,d_1^*,1,\cdots,N,N^*,\cdots,1^*), \\
&\tau_{xxx}=pf(d_2,d_0^*,1,\cdots,N,N^*,\cdots,1^*)+2pf(d_1,d_1^*,1,\cdots,N,N^*,\cdots,1^*) \\
&\qquad +pf(d_0,d_2^*,1,\cdots,N,N^*,\cdots,1^*) \\
&\tau_{y}=-pf(d_1,d_0^*,1,\cdots,N,N^*,\cdots,1^*) +pf(d_0,d_1^*,1,\cdots,N,N^*,\cdots,1^*),  \\
&\tau_{yx}=-pf(d_2,d_0^*,1,\cdots,N,N^*,\cdots,1^*) +pf(d_0,d_2^*,1,\cdots,N,N^*,\cdots,1^*), \\
&\tau_t=pf(d_2,d_0^*,1,\cdots,N,N^*,\cdots,1^*)-pf(d_1,d_1^*,1,\cdots,N,N^*,\cdots,1^*) \\
 &\qquad \, +pf(d_0,d_2^*,1,\cdots,N,N^*,\cdots,1^*)+\sum_{i=1} ^N \dot c_i pf(1,\cdots,\hat{i},\cdots,N,N^*,\cdots,\hat{i^*},\cdots,1^*) \\
&\tau'_x=pf( d_{1},d_{-1}^*,1,\cdots,N,N^*,\cdots,1^*), \\
&\tau'_{xx}= pf( d_{2},d_{-1}^*,1,\cdots,N,N^*,\cdots,1^*) + pf( d_{1},d_{0}^*,1,\cdots,N,N^*,\cdots,1^*)m \\
&\qquad+pf( d_{0},d_{0}^*,d_{1},d_{-1}^*,1,\cdots,N,N^*,\cdots,1^*), \\
&\tau'_{xxx}=pf(d_3,d_{-1}^*,1,\cdots,N,N^*,\cdots,1^*) +2pf(d_2,d_0^*,1,\cdots,N,N^*,\cdots,1^*) \\& \qquad\,\,+pf(d_1,d_1^*,1,\cdots,N,N^*,\cdots,1^*) 
+pf( d_{0},d_{1}^*,d_{1},d_{-1}^*,1,\cdots,N,N^*,\cdots,1^*) \\ 
&\qquad\,\,+pf( d_{0},d_{0}^*,d_{2},d_{-1}^*,1,\cdots,N,N^*,\cdots,1^*), \\
&\tau'_{y}=- pf( d_{2},d_{-1}^*,1,\cdots,N,N^*,\cdots,1^*) + pf( d_{1},d_{0}^*,1,\cdots,N,N^*,\cdots,1^*) \\
& \qquad-pf( d_{0},d_{-1}^*,d_{1},d_{0}^*,1,\cdots,N,N^*,\cdots,1^*), \\
&\tau'_{xy}=-pf(d_3,d_{-1}^*,1,\cdots,N,N^*,\cdots,1^*) +pf(d_1,d_1^*,1,\cdots,N,N^*,\cdots,1^*) \\
&\qquad\,\,-pf( d_{0},d_{-1}^*,d_{1},d_{1}^*,1,\cdots,N,N^*,\cdots,1^*), \\
 & \tau'_t=pf(d_0,d_2^*,1,\cdots,N,N^*,\cdots,1^*) -pf(d_1,d_1^*,1,\cdots,N,N^*,\cdots,1^*) \\
 & \qquad \, -pf(d_{-1},d_3^*,1,\cdots,N,N^*,\cdots,1^*)-pf(d_{-1},d_0^*,d_1,d_1^*,1,\cdots,N,N^*,\cdots,1^*) \\
 & \qquad \, -pf(d_{-1},d_0^*,d_0,d_2^*,1,\cdots,N,N^*,\cdots,1^*)+\sum_{i=1} ^N \dot c_i pf(1,\cdots,\hat{i},\cdots,N,N^*,\cdots,\hat{i^*},\cdots,1^*) \\
 &\qquad \, -\sum_{i=1} ^N \dot c_i pf(d_{-1},d_0^*,1,\cdots,\hat{i},\cdots,N,N^*,\cdots,\hat{i^*},\cdots,1^*) \\
&\Phi_{i,x}=\left[ pf(d_1^*,1,\cdots,N,N^*,\cdots,\hat{i^*},\cdots,1^*) \right] \beta_i,
\\
&\Phi_{i,y}= \left[ pf(d_2^*,1,\cdots,N,N^*,\cdots,\hat{i^*},\cdots,1^*) -pf(d_0,d_{0}^*,d_1^*,1,\cdots,N,N^*,\cdots,\hat{i^*},\cdots,1^*)\right] \beta_i,\\
&\Phi_{i,xx}= \left[ pf(d_2^*,1,\cdots,N,N^*,\cdots,\hat{i^*},\cdots,1^*)+pf(d_0,d_{0}^*,d_1^*,1,\cdots,N,N^*,\cdots,\hat{i^*},\cdots,1^*)\right] \beta_i, \\
&\Psi_{i,x}=-  \eta_i [pf(d_{0},1,\cdots,\hat{i},\cdots,N,N^*,\cdots,1^*)+pf(d_{-1},d_0,d_0^*,1,\cdots,\hat{i},\cdots,N,N^*,\cdots,1^*)], \\
&\Psi_{i,xx}= - \eta_i [ pf(d_1,1,\cdots,\hat{i},\cdots,N,N^*,\cdots,1^*) +pf(d_{-1},d_1,d_0^*,1,\cdots,\hat{i},\cdots,N,N^*,\cdots,1^*) \\
&\qquad\quad\,+pf(d_{-1},d_0,d_1^*,1,\cdots,\hat{i},\cdots,N,N^*,\cdots,1^*) ],  \\ 
&\Psi_{i,y}=  \eta_i [ pf(d_1,1,\cdots,\hat{i},\cdots,N,N^*,\cdots,1^*) +pf(d_{-1},d_1,d_0^*,1,\cdots,\hat{i},\cdots,N,N^*,\cdots,1^*) \\
&\qquad\quad-pf(d_{-1},d_0,d_1^*,1,\cdots,\hat{i},\cdots,N,N^*,\cdots,1^*) ], 
\end{align*}


\begin{thebibliography}{9}



\bibitem{sineGorden2002}
M Grisaru and S Penati 2002   {\it Nuclear Physics B } {\bf 655} 250

\bibitem{Hamanaka_2003}
M Hamanaka and K Toda 2003 {\it Physics Letters A } {\bf 316} 77

\bibitem{2003Hamanaka_}
M Hamanaka and K Toda 2003 {\it  Journal of Physics A: Mathematical and General } {\bf 36} 11981

\bibitem{2005GELFAND}
I Gelfand, S Gelfand, V Retakh and R Wilson 2005 {\it Advances in Mathematics} {\bf 193} 56

\bibitem{2007Gilson} 
C R Gilson J J C Nimmo and Y Ohta 2007 {\it Journal of Physics A: Mathematical and Theoretical}
{\bf 40} 12607

\bibitem{2008Li_} 
C X Li and J J C Nimmo 2008 {\it Proceedings of the Royal Society A: Mathematical, Physical and
Engineering Sciences} {\bf 464} 951

\bibitem{2007non-Abelian}
C R Gilson and J J C Nimmo 2007  {\it Journal of Physics A: Mathematical and Theoretical}  {\bf 40} 3839

\bibitem{2008Gilson}
C R Gilson, J J C Nimmo, and C M Sooman 2008 {\it Journal of Physics A: Mathematical
and Theoretical}  {\bf 41 } 85202

\bibitem{2009Li}
C X Li, J J C Nimmo and K M Tamizhmani 2009 {\it Proceedings of the Royal Society A: Mathematical,
Physical and Engineering Sciences}  {\bf 465} 1441

\bibitem{2009Gilson_}
C R Gilson, J J C Nimmo and C M Sooman 2009  {\it Theoretical and Mathematical Physics}  {\bf 159} 
 796

\bibitem{2010Wu}
W X Ma 2010  {\it Journal of Mathematical Physics} {\bf 51} 073505

\bibitem{2017Li}
H Wu, J Liu, and C X Li. 2017   {\it Theoretical and Mathematical Physics}  {\bf 192} 982

\bibitem{2008Wang}
 H Wang, X B Hu and H W Tam 2008 {\it Journal of Mathematical Analysis and
Applications} {\bf 338} 82

\bibitem{2005Xiao_}
T Xiao and Y Zeng 2005  {\it Journal of Physics A: Mathematical and General}  {\bf 39} 139

\bibitem{2009Liu}
X Liu, R Lin, B Jin and Y Zeng 2009 {\it Journal of Mathematical Physics}  {\bf 50} 053506

\bibitem{2005XIAO}
T Xiao and Y Zeng 2005  {\it  Physica A: Statistical Mechanics and its Applications}  {\bf 353} 38

\bibitem{2008LIU}
X Liu, R Lin, B Jin and Y Zeng 2009 {\it Journal of Mathematical Physics} {\bf 50} 053506

\bibitem{1991Gelfand}
I Gel’fand and V Retakh 1991  {\it Functional Analysis and Its Applications}  {\bf 25} 91

\bibitem{1989CMaPh.126..201M}
V K Mel’nikov 1989 {\it Communications in Mathematical Physics}  {\bf  126} 201

\bibitem{2004Xiao}
T Xiao and Y Zeng 2004 {\it Journal of Physics A: Mathematical and General}  {\bf 37} 7143

\bibitem{2003Shu-fang}
C Deng-Yuan and D Shu-fang and Z Da-Jun 2003 {\it  Journal of the Physical Society of Japan}  {\bf 72} 2184

\bibitem{2004hirota}
R Hirota 2004 The Direct Method in Soliton Theory {\it Cambridge University Press}

\bibitem{2006Dimakis_}
A Dimakis and  F M Hoissen 2006 {\it  Journal of Physics A: Mathematical and Gen-
eral}  {\bf 39} 9169
\end{thebibliography}
\end{document}